\documentclass[prd,preprint,superscriptaddress,showpacs,byrevtex]{revtex4}
\usepackage{bm}
\def\fsl#1{\setbox0=\hbox{$#1$}           % set a box for #1
   \dimen0=\wd0                                 % and get its size
   \setbox1=\hbox{/} \dimen1=\wd1               % get size of /
   \ifdim\dimen0>\dimen1                        % #1 is bigger
      \rlap{\hbox to \dimen0{\hfil/\hfil}}      % so center / in box
      #1                                        % and print #1
   \else                                        % / is bigger
      \rlap{\hbox to \dimen1{\hfil$#1$\hfil}}   % so center #1
      /                                         % and print /
   \fi}                                         %
\newcommand{\be}{\begin{equation}}
\newcommand{\ee}{\end{equation}}
\newcommand{\bea}{\begin{eqnarray}}
\newcommand{\eea}{\end{eqnarray}}
\newcommand{\beq}{\begin{equation}}
\newcommand{\eeq}{\end{equation}}
\newcommand{\beqs}{\begin{eqnarray}}
\newcommand{\eeqs}{\end{eqnarray}}

\newcommand{\aslash}{A\hspace{-0.067in}\slash}

\begin{document}
\title{ $\Upsilon$ Production From Quark-Gluon Plasma }
\author{Gouranga C Nayak }\thanks{E-Mail: nayakg138@gmail.com}
%
%\affiliation{ C. N. Yang Institute for Theoretical Physics, Stony Brook University, Stony Brook NY, 11794-3840 USA}
%
\date{\today}
\begin{abstract}
Complete suppression of the heavy quarkonium due to the screening mechanism of the quark-gluon plasma is proposed in the literature to be a signature of the quark-gluon plasma detection at RHIC and LHC. However, since the heavy quarkonium $\Upsilon$ production is experimentally measured in the Pb-Pb collisions by the ALICE collaboration at LHC one has to study the $\Upsilon$ production from the quark-gluon plasma instead of studying the $\Upsilon$ suppression due to the screening mechanism of the quark-gluon plasma. In this paper we derive the non-perturbative formula of the $\Upsilon$ production amplitude from the quark-gluon plasma from the first principle in QCD which can be calculated by using the lattice QCD method at the finite temperature.
\end{abstract}
\maketitle
\pagestyle{plain}

\pagenumbering{arabic}

\section{ Introduction }

Just after $10^{-12}$ seconds of the big bang our universe was filled with a hot and dense state of matter known as the quark-gluon plasma (QGP). The temperature of the quark-gluon plasma is $\sim $ 200 MeV ($\sim 10^{12}$ K) which is about million times larger than the temperature of the sun. The quark-gluon plasma is also the densest state of the matter of the universe just after the black hole.

The hadrons (such as proton and neutron) were formed for the first time in the universe from this quark-gluon plasma. Note that almost all of the visible matter of the universe are made from the protons, neutrons and electrons. Hence it is important to recreate this early universe scenario in the laboratory, {\it i. e.}, it is important to produce the quark-gluon plasma in the laboratory.

The two experiments in the laboratory which study the production of quark-gluon plasma are: 1) RHIC (relativistic heavy-ion colliders) and 2) LHC (large hadron colliders) \cite{qg,qg1,qg2,qg3}. The RHIC (LHC) experiment involves the Au-Au (Pb-Pb) collisions at the center of mass energy per nucleon ${\sqrt s}_{NN}$ = 200 GeV (5.02 TeV).

The quantum chromodynamics (QCD) \cite{ym} is the fundamental theory of the nature which describes the interaction between quarks and gluons which are the fundamental particles of the nature. Due to the asymptotic freedom in QCD \cite{gw} the perturbative QCD (pQCD) can be applied to calculate the short distance (high momentum transfer) cross section at the high energy colliders. Using the factorization theorem in QCD \cite{fc,fc1,fc2} the hadronic cross section at the high energy colliders is calculated from the partonic cross section by using the experimentally extracted parton distribution function (PDF) and the fragmentation function (FF).

The hadron formation from quarks and gluons is a long distance phenomenon in QCD which cannot be studied by using the pQCD. The non-perturbative QCD is necessary for this purpose. Since the analytical solution of the non-perturbative QCD is not known yet the lattice QCD method can be used to study the hadron formation from the quarks and gluons.

Note that we have not directly experimentally observed the quarks and gluons. What we experimentally observe are the hadrons. Hence we cannot directly detect the quark-gluon plasma at RHIC and LHC. Due to this reason various indirect signatures are proposed for the quark-gluon plasma detection at RHIC and LHC. The indirect hadronic signatures for the quark-gluon plasma detection at RHIC and LHC are: 1) jet quenching, 2) strangeness enhancement and 3) heavy quarkonium suppression. We will focus on the heavy quarkonium suppression/production signature in this paper.

The heavy quarkonium suppression as a signature of quark-gluon plasma detection was based on the idea of Debye screening by Matsui and Satz \cite{ms}. They found that the Debye screening length in the high temperature quark-gluon plasma is less than the heavy quakonium radius preventing the heavy quarkonium formation. According to this screening idea of the quark-gluon plasma the heavy quarkonium is completely suppressed.

However, the heavy quarkonium $\Upsilon$ production is experimentally measured in the Pb-Pb collisions at $\sqrt{s}=$ 5.02 TeV by the ALICE collaboration at the LHC \cite{alice}. This implies that the $\Upsilon$ is not completely suppressed due to the presence of the quark-gluon plasma as suggested by the screening mechanism. Hence one has to study the $\Upsilon$ production from the quark-gluon plasma instead of studying the $\Upsilon$ suppression due to the screening mechanism of the quark-gluon plasma.

In this paper we derive the non-perturbative formula of the $\Upsilon$ production amplitude from the quark-gluon plasma from the first principle in QCD which can be calculated by using the lattice QCD method at the finite temperature.

The paper is organized as follows. In section II we describe the $\Upsilon$ formation from the quarks and gluons in QCD in vacuum using the lattice QCD method. In section III we derive the non-perturbative formula of the $\Upsilon$ production amplitude from the quark-gluon plasma from the first principle in QCD which can be calculated by using the lattice QCD method at the finite temperature. Section IV contains conclusions.

\section{ $\Upsilon$ Formation From Quarks and Gluons in QCD in vacuum Using lattice QCD method}

The generating functional in QCD is given by
\bea
&& Z[0] =\int [d{\bar \Psi}_U [d\Psi_U] [d{\bar \Psi}_D] [d\Psi_D] [d{\bar \Psi}_S] [d\Psi_S] [d{\bar \Psi}_C] [d\Psi_C] [d{\bar \Psi}_B] [d\Psi_B][dA] \times {\rm det}[\frac{\delta {\cal G}_{\cal F}^c}{\delta \omega^a} ]\nonumber \\
&& \times {\rm exp}[i\int d^4x [-\frac{1}{4} F_{\lambda \mu}^b(x) F^{\lambda \mu b}(x) -\frac{1}{2\alpha}[{\cal G}_{\cal F}^b(x)]^2 +{\bar \Psi}_U^j(x)[\delta^{jl}(i{\not \partial}-M_U)+gT^b_{jl}\aslash^b(x)]\Psi_U^l(x) \nonumber \\
&& +{\bar \Psi}_D^j(x)[\delta^{jl}(i{\not \partial}-M_D)+gT^b_{jl}\aslash^b(x)]\Psi_D^l(x)+{\bar \Psi}_S^j(x)[\delta^{jl}(i{\not \partial}-M_S)+gT^b_{jl}\aslash^b(x)]\Psi_S^l(x)\nonumber \\
&&+{\bar \Psi}_C^j(x)[\delta^{jl}(i{\not \partial}-M_C)+gT^b_{jl}\aslash^b(x)]\Psi_C^l(x)+{\bar \Psi}_B^j(x)[\delta^{jl}(i{\not \partial}-M_B)+gT^b_{jl}\aslash^b(x)]\Psi_B^l(x)
\label{gf}
\eea
where $\Psi_f^K(x)$ is the field of the quark of flavor $f=U,D,S,C,B$ (= up, down, strange, charm, bottom), the $k=1,2,3$ is the color index, ${\cal G}_{\cal F}^a(x)$ is the gauge fixing term, $\alpha$ is the gauge fixing parameter, $A_\mu^a(x)$ is the gluon field with color index $a=1,...,8$ and Lorentz index $\mu=0.1.2.3$.
In eq. (\ref{gf})
\bea
F_{\lambda \mu}^b(x) =\partial_\lambda A_\nu^b(x) - \partial_\lambda A_\nu^b(x) +gf^{bsc} A_\lambda^s(x) A_\nu^c(x)
\label{fb}
\eea
and we do not have ghost fields because we directly work with the ghost determinant ${\rm det}[\frac{\delta {\cal G}_{\cal F}^c}{\delta \omega^a} ]$.

The partonic operator for the heavy quarkonium $\Upsilon$ formation is given by
\bea
{\cal O}_\Upsilon(x)= {\bar \Psi}_B^k(x) \gamma^\mu \Psi_B^k(x).
\label{sg}
\eea

The energy-momentum tensor density $T^{\lambda \mu}(x)$ in QCD is given by
\bea
&&T^{\lambda \mu}(x)=
F^{\lambda \delta b}(x)F_{\delta }^{~\mu b}(x)+\frac{g^{\lambda \mu}}{4} F^{\delta \delta' b}(x)F_{\delta \delta'}^b(x)+{\bar \Psi}_U^j(x) \gamma^\lambda [\delta^{jk}i \partial^{\mu} -i gT^b_{jk}A^{b \mu}(x)]\Psi_U^k(x)\nonumber \\
&&+{\bar \Psi}_D^j(x) \gamma^\lambda [\delta^{jk}i \partial^{\mu} -i gT^b_{jk}A^{b \mu}(x)]\Psi_D^k(x)+{\bar \Psi}_S^j(x) \gamma^\lambda [\delta^{jk}i \partial^{\mu} -i gT^b_{jk}A^{b \mu}(x)]\Psi_S^k(x)\nonumber \\
&&+{\bar \Psi}_C^j(x) \gamma^\lambda [\delta^{jk}i \partial^{\mu} -i gT^b_{jk}A^{b \mu}(x)]\Psi_C^k(x)+{\bar \Psi}_B^j(x) \gamma^\lambda [\delta^{jk}i \partial^{\mu} -i gT^b_{jk}A^{b \mu}(x)]\Psi_B^k(x).
\label{gtp}
\eea
From the continuity equation
\bea
\partial_\lambda T^{\lambda \mu}(x)=0
\label{cte}
\eea
we find
\bea
\frac{dE(p,t)}{dt}=-\frac{dE_S(p,t)}{dt}\neq 0
\label{dtp1}
\eea
where $E(p,t)$ is the energy of all the partons inside the bottomonium $\Upsilon$ of momentum $p^\mu$ given by
\bea
E(p,t)=<\Upsilon(p)|\int d^3x \sum_{q,{\bar q},g} T^{00}(t,{\bf x})|\Upsilon(p)>
\label{etp}
\eea
and $E_S(p,t)$ is the non-zero boundary term in QCD due to the confinement of quarks and gluons inside the finite size $\Upsilon$ given by [see eq. (\ref{netf2})]
\bea
\frac{dE_S(p,t)}{dt}=<\Upsilon(p)|\int d^3x \sum_{q,{\bar q},g} \partial_j T^{j0}(t,{\bf x})|\Upsilon(p)>\neq 0.
\label{ftp}
\eea
In eqs. (\ref{etp}) and (\ref{ftp}) the $|\Upsilon(p)>$ is the energy-momentum eigenstate of the heavy quarkonium $\Upsilon$ of momentum $p^\mu$.

The non-zero boundary term in eq. (\ref{ftp}) can be seen as follows. We know that the quark and gluon fields carry color but we do not observe color outside the hadron because the hadron is colorless. This implies that the quark and gluon fields do not exist outside the hadron. The volume of the hadron is finite. Hence the volume integral $\int d^3x$ in $<\Upsilon({\vec p})|\sum_{q,{\bar q},g} \int d^3x \partial_i T^{i 0}_{q{\bar q}g}(t,{\bf x})|\Upsilon({\vec p})>$ in eq. (\ref{ftp}) encloses the volume of the finite size hadron containing the quark and gluon fields which gives non-vanishing boundary term in eq. (\ref{ftp}).

One may argue that one may stop at distances much larger than the hadron size where the quark and gluon fields are zero to obtain vanishing boundary term. However, this argument is not correct which can be seen as follows.

Suppose one stops at distances much larger than the hadron size, say within a total volume
\bea
\int d^3x=\int _{\le R} d^3x+\int_{>R}d^3x
\label{vl}
\eea
where $\int_{\le R}d^3x$ is the volume of the hadron which contains the non-zero quark and gluon fields and $\int_{>R}d^3x$ is the remaining volume outside the hadron where the quark and gluon fields are zero. Since the quark and gluon fields are zero in the volume $\int_{>R}d^3x$ we find
\bea
&& <\Upsilon({\vec p})|\sum_{q,{\bar q},g} \int_{{>R}} d^3x \partial_i T^{i 0}_{q{\bar q}g}(t,{\bf x})|\Upsilon({\vec p})> =0.
\label{netf2a}
\eea
The boundary of $\int d^3x $ in eq. (\ref{vl}) is at a larger distance than the boundary of $\int_{\le R}d^3x$, {\it i. e.}, the boundary of $\int d^3x$ in eq. (\ref{vl}) is at a larger distance than the boundary of the hadron. Since the quark and gluon fields are non-zero inside the hadron and zero outside the hadron one finds that the quark and gluon fields inside the volume $\int d^3x$ in eq. (\ref{vl}) are not continuously differentiable near the boundary of the hadron. Hence one can not apply the divergence theorem to such fields near the boundary of $\int d^3x$ in eq. (\ref{vl}) because the divergence theorem is applicable to continuously differentiable function inside $\int d^3x$. Since one cannot apply the divergence theorem at the boundary of $\int d^3x$ one cannot prove that the boundary term is zero. In this situation where the fields are not continuously differentiable inside the volume $\int d^3x$ one has to perform the volume integration as follows.

In the volume $\int d^3x$, since the fields are not continuously differentiable near the boundary of the hadron, one has to separate the volume integral $\int d^3x$ into two parts as given by eq. (\ref{vl}) where $\int_{\le R}d^3x$ is the volume of the hadron which contains the non-zero quark and gluon fields and $\int_{>R}d^3x$ is the remaining volume outside the hadron where the quark and gluon fields are zero. As mentioned above since the divergence theorem is not applicable to fields which are not continuously differentiable inside the volume $\int d^3x$ one has to directly evaluate the volume integral to find
\bea
&& <\Upsilon({\vec p})|\sum_{q,{\bar q},g} \int d^3x \partial_i T^{i 0}_{q{\bar q}g}(t,{\bf x})|\Upsilon({\vec p})> \nonumber \\
&&=<\Upsilon({\vec p})|\sum_{q,{\bar q},g} \int_{{\le R}} d^3x \partial_i T^{i 0}_{q{\bar q}g}(t,{\bf x})|\Upsilon({\vec p})> +<\Upsilon({\vec p})|\sum_{q,{\bar q},g} \int_{{>R}} d^3x \partial_i T^{i 0}_{q{\bar q}g}(t,{\bf x})|\Upsilon({\vec p})> \nonumber \\
&&=<\Upsilon({\vec p})|\sum_{q,{\bar q},g} \int_{{\le R}} d^3x \partial_i T^{i0}_{q{\bar q}g}(t,{\bf x})|\Upsilon({\vec p})> \neq 0
\label{netf2}
\eea
which proves eq. (\ref{ftp}) where we have used eq. (\ref{netf2a}) in the second line of eq. (\ref{netf2}).

The time evolution of the operator ${\cal O}_\Upsilon(t,{\bf x})$ is given by
\bea
{\cal O}_\Upsilon(t,{\bf x}) = e^{-iHt} {\cal O}_\Upsilon(0,{\bf x}) e^{iHt}
\label{tev}
\eea
where $H$ means the QCD hamiltonian. The complete set of energy-momentum eigenstates $|\Upsilon_n(p)>$ of the heavy quarkonium $\Upsilon$ of momentum $p^\mu$ is given by
\bea
\sum_n|\Upsilon_n(p)><\Upsilon_n(p)|=1.
\label{egs}
\eea
The two-point non-perturbative correlation function of the type $<0|{\cal O}_\Upsilon^\dagger(t',{\bf x}'){\cal O}_\Upsilon(0)|0>$ in QCD is given by
\bea
&&<0|{\cal O}_\Upsilon^\dagger(t',{\bf x}'){\cal O}_\Upsilon(0)|0>=\frac{1}{Z[0]}\int [d{\bar \Psi}_U [d\Psi_U] [d{\bar \Psi}_D] [d\Psi_D] [d{\bar \Psi}_S] [d\Psi_S] [d{\bar \Psi}_C] [d\Psi_C] [d{\bar \Psi}_B] [d\Psi_B][dA]\nonumber \\
&& \times {\cal O}_\Upsilon^\dagger(t',{\bf x}'){\cal O}_\Upsilon(0)\nonumber \times {\rm det}[\frac{\delta {\cal G}_{\cal F}^c}{\delta \omega^a} ]\nonumber \\
&&\times {\rm exp}[i\int d^4x [-\frac{1}{4} F_{\lambda \mu}^b(x) F^{\lambda \mu b}(x) -\frac{1}{2\alpha}[{\cal G}_{\cal F}^b(x)]^2 +{\bar \Psi}_U^j(x)[\delta^{jl}(i{\not \partial}-M_U)+gT^b_{jl}\aslash^b(x)]\Psi_U^l(x) \nonumber \\
&& +{\bar \Psi}_D^j(x)[\delta^{jl}(i{\not \partial}-M_D)+gT^b_{jl}\aslash^b(x)]\Psi_D^l(x)+{\bar \Psi}_S^j(x)[\delta^{jl}(i{\not \partial}-M_S)+gT^b_{jl}\aslash^b(x)]\Psi_S^l(x)\nonumber \\
&&+{\bar \Psi}_C^j(x)[\delta^{jl}(i{\not \partial}-M_C)+gT^b_{jl}\aslash^b(x)]\Psi_C^l(x)+{\bar \Psi}_B^j(x)[\delta^{jl}(i{\not \partial}-M_B)+gT^b_{jl}\aslash^b(x)]\Psi_B^l(x)
\label{tp}
\eea
where $|0>$ is the vacuum state of the non-perturbative QCD (not the vacuum state of the pQCD).

Using eqs. (\ref{tev}) and (\ref{egs}) in eq. (\ref{tp}) we find in the Euclidean time
\bea
&&\sum_{{\bf x}'} e^{i{\bf p}\cdot {\bf x}'} <0|{\cal O}_\Upsilon^\dagger(t',{\bf x}'){\cal O}_\Upsilon(0)|0>=\sum_l |<\Upsilon_l(p)|{\cal O}_\Upsilon|0>|^2 e^{-\int dt' E_l(p,t')}
\label{atp}
\eea
where $\int dt'$ is an indefinite integration, $E_l(p,t)$ is the energy of all the quarks and gluons inside the bottomonium $\Upsilon$ of momentum $p^\mu$ in its l$th$ energy level state.

Neglecting the higher energy level contributions in the asymptotic large time limit $t'\rightarrow \infty$ we find
\bea
&&[\sum_{{\bf x}'} e^{i{\bf p}\cdot {\bf x}'} <0|{\cal O}_\Upsilon^\dagger(t',{\bf x}'){\cal O}_\Upsilon(0)|0>]_{t'\rightarrow \infty} = |<\Upsilon(p)|{\cal O}_\Upsilon|0>|^2 e^{-\int dt' E(p,t')}
\label{btp}
\eea
where
\bea
|\Upsilon(p)>=|\Upsilon(p)>,~~~~~~~~~~~~~~~E_0(p,t)=E(p,t).
\label{ctp}
\eea
From eqs. (\ref{dtp1}) and (\ref{ftp}) we find
\bea
\frac{d}{dt}[E(p,t)+E_S(p,t)]=0
\label{dtp2}
\eea
which gives
\bea
E_\Upsilon(p) = E(p,t)+E_S(p,t)
\label{dtp}
\eea
where $E_\Upsilon(p)$ is the conserved [time independent] energy of the bottomonium $\Upsilon$ of momentum $p^\mu$, the $E(p,t)$ is the energy of all the partons inside the bottomonium $\Upsilon$ of momentum $p^\mu$ given by eq. (\ref{etp}) and $E_S(p,t)$ is the non-zero boundary term in QCD due to the confinement of quarks and gluons inside the finite size $\Upsilon$ given by eq. (\ref{ftp}).

As mentioned above since we do not directly experimentally observe the quarks and gluons we do not directly experimentally observe the energy $E(p,t)$ of the partons  given by eq. (\ref{etp}) and we do not directly experimentally observe the non-zero boundary term (the non-zero energy flow) $E_S(p,t)$ of the partons in QCD in eq. (\ref{ftp}) but we directly experimentally observe the energy $E_\Upsilon(p)$ of the hadron $\Upsilon$ in eq. (\ref{dtp}). Hence the the non-zero boundary term (the non-zero energy flow) $E_S(p,t)$ of the partons in QCD contributes to the energy $E_\Upsilon(p)$ of the hadron $\Upsilon$ which we experimentally observe.

Similar to the derivation of eq. (\ref{btp}) we find from eq. (\ref{ftp}) that
\bea
&&\frac{dE_S(p,t)}{dt}=[\frac{\sum_{{\bf x}'} e^{i{\bf p}\cdot {\bf x}'} <0|{\cal O}_\Upsilon^\dagger(t',{\bf x}')[\sum_{q,{\bar q},g} \int d^3x \partial_j T^{j0}(t,{\bf x})]{\cal O}_\Upsilon(0)|0>}{\sum_{{\bf x}'} e^{i{\bf p}\cdot {\bf x}'} <0|{\cal O}_\Upsilon^\dagger(t',{\bf x}'){\cal O}_\Upsilon(0)|0>}]_{t'\rightarrow \infty}.
\label{htp}
\eea
Using eqs. (\ref{dtp}) and (\ref{htp}) in eq. (\ref{btp}) we find
\bea
&&|<\Upsilon(p)|{\cal O}_\Upsilon|0>|^2 =[\frac{\sum_{{\bf x}} e^{i{\bf p}\cdot {\bf x}} <0|{\cal O}_\Upsilon^\dagger(t,{\bf x}){\cal O}_\Upsilon(0)|0>e^{t E_\Upsilon(p)}}{e^{\int dt [\frac{\sum_{{\bf x}'} e^{i{\bf p}\cdot {\bf x}'} <0|{\cal O}_\Upsilon^\dagger(t',{\bf x}')[\sum_{q,{\bar q},g}\int dt \int d^3x  \partial_j T^{j0}(t,{\bf x})]{\cal O}_\Upsilon(0)|0>}{\sum_{{\bf x}'} e^{i{\bf p}\cdot {\bf x}'} <0|{\cal O}_\Upsilon^\dagger(t',{\bf x}'){\cal O}_\Upsilon(0)|0>}]_{t'\rightarrow \infty}}} ]_{t\rightarrow \infty}
\label{itp}
\eea
which is the non-perturbative formula to study the bottomonium $\Upsilon$ formation from the quarks and gluons in QCD in vacuum from the first principle which can be calculated by using the lattice QCD method.

\section{ $\Upsilon$ Production From quark-gluon plasma Using Lattice QCD Method at Finite Temperature }

The generating functional in QCD at the finite temperature $T$ is given by
\bea
&& Z[0] =\int [d{\bar \Psi}_U [d\Psi_U] [d{\bar \Psi}_D] [d\Psi_D] [d{\bar \Psi}_S] [d\Psi_S] [d{\bar \Psi}_C] [d\Psi_C] [d{\bar \Psi}_B] [d\Psi_B][dA] \times {\rm det}[\frac{\delta {\cal G}_{\cal F}^c}{\delta \omega^a} ] \times {\rm exp}[-\int_0^{1/T} d\tau \nonumber \\
&& \int d^3x [-\frac{1}{4} F_{\lambda \mu}^b(\tau,{\bf x}) F^{\lambda \mu b}(\tau,{\bf x}) -\frac{1}{2\alpha}[{\cal G}_{\cal F}^b(\tau,{\bf x})]^2 +{\bar \Psi}_U^j(\tau,{\bf x})[\delta^{jl}(i{\not \partial}-M_U)+gT^b_{jl}\aslash^b(\tau,{\bf x})]\Psi_U^l(\tau,{\bf x}) \nonumber \\
&& +{\bar \Psi}_D^j(\tau,{\bf x})[\delta^{jl}(i{\not \partial}-M_D)+gT^b_{jl}\aslash^b(\tau,{\bf x})]\Psi_D^l(\tau,{\bf x})+{\bar \Psi}_S^j(\tau,{\bf x})[\delta^{jl}(i{\not \partial}-M_S)+gT^b_{jl}\aslash^b(\tau,{\bf x})]\Psi_S^l(\tau,{\bf x})\nonumber \\
&&+{\bar \Psi}_C^j(\tau,{\bf x})[\delta^{jl}(i{\not \partial}-M_C)+gT^b_{jl}\aslash^b(\tau,{\bf x})]\Psi_C^l(\tau,{\bf x})+{\bar \Psi}_B^j(\tau,{\bf x})[\delta^{jl}(i{\not \partial}-M_B)+gT^b_{jl}\aslash^b(\tau,{\bf x})]\Psi_B^l(\tau,{\bf x})\nonumber \\
\label{tgf}
\eea
where the quark field satisfies the anti-periodic boundary condition and the gluon field satisfies the periodic boundary condition
\bea
\Psi^k(\tau,{\bf x})=-\Psi^k(\tau+\frac{1}{T},{\bf x}),~~~~~~~~~~~~  A_\lambda^b(\tau,{\bf x})=A_\lambda^b(\tau+\frac{1}{T},{\bf x}).
\label{tpdc}
\eea
The two-point non-perturbative correlation function of the type $<in|{\cal O}_\Upsilon^\dagger(\tau',{\bf x}'){\cal O}_\Upsilon(0)|in>$ in QCD at the finite temperature $T$ is given by
\bea
&&<in|{\cal O}_\Upsilon^\dagger(\tau',{\bf x}'){\cal O}_\Upsilon(0)|in>=\frac{1}{Z[0]}\int [d{\bar \Psi}_U [d\Psi_U] [d{\bar \Psi}_D] [d\Psi_D] [d{\bar \Psi}_S] [d\Psi_S] [d{\bar \Psi}_C] [d\Psi_C] [d{\bar \Psi}_B] [d\Psi_B][dA]\nonumber \\
&& \times {\cal O}_\Upsilon^\dagger(\tau',{\bf x}'){\cal O}_\Upsilon(0)\nonumber \times {\rm det}[\frac{\delta {\cal G}_{\cal F}^c}{\delta \omega^a} ]\times {\rm exp}[-\int_0^{1/T} d\tau \int d^3x [-\frac{1}{4} F_{\lambda \mu}^b(\tau,{\bf x}) F^{\lambda \mu b}(\tau,{\bf x}) -\frac{1}{2\alpha}[{\cal G}_{\cal F}^b(\tau,{\bf x})]^2\nonumber \\
&&  +{\bar \Psi}_U^j(\tau,{\bf x})[\delta^{jl}(i{\not \partial}-M_U)+gT^b_{jl}\aslash^b(\tau,{\bf x})]\Psi_U^l(\tau,{\bf x}) +{\bar \Psi}_D^j(\tau,{\bf x})[\delta^{jl}(i{\not \partial}-M_D)+gT^b_{jl}\aslash^b(\tau,{\bf x})]\Psi_D^l(\tau,{\bf x})\nonumber \\
&& +{\bar \Psi}_S^j(\tau,{\bf x})[\delta^{jl}(i{\not \partial}-M_S)+gT^b_{jl}\aslash^b(\tau,{\bf x})]\Psi_S^l(\tau,{\bf x})+{\bar \Psi}_C^j(\tau,{\bf x})[\delta^{jl}(i{\not \partial}-M_C)+gT^b_{jl}\aslash^b(\tau,{\bf x})]\Psi_C^l(\tau,{\bf x})\nonumber \\
&& +{\bar \Psi}_B^j(\tau,{\bf x})[\delta^{jl}(i{\not \partial}-M_B)+gT^b_{jl}\aslash^b(\tau,{\bf x})]\Psi_B^l(\tau,{\bf x})
\label{ttp}
\eea
where $|in>$ is the ground state of the non-perturbative QCD at the finite temperature $T$ (not the ground state of the pQCD at the finite temperature $T$).

Unlike
\bea
<0|H=0
\eea
in QCD in vacuum, since
\bea
<in|H\neq 0
\eea
in the finite temperature QCD we find by using eqs. (\ref{tev}) and (\ref{egs}) in eq. (\ref{ttp}) in the Euclidean time that
\bea
&&\sum_{{\bf x}'} e^{i{\bf p}\cdot {\bf x}'} <in|e^{-H\tau'} {\cal O}_\Upsilon^\dagger(\tau',{\bf x}'){\cal O}_\Upsilon(0)|in>=\sum_l |<\Upsilon_l(p)|{\cal O}_\Upsilon|in>|^2 e^{-\int d\tau' E_l(p,\tau')}
\label{tatpz}
\eea
where $\int d\tau'$ is an indefinite integration. Using eq. (\ref{tev}) in (\ref{tatpz}) we find
\bea
&&\sum_{{\bf x}'} e^{i{\bf p}\cdot {\bf x}'} <in|{\cal O}_\Upsilon^\dagger(0,{\bf x}')e^{-H\tau'}{\cal O}_\Upsilon(0)|in>=\sum_l |<\Upsilon_l(p)|{\cal O}_\Upsilon|in>|^2 e^{-\int d\tau' E_l(p,\tau')}.
\label{tatp}
\eea

Note that the bottomonium $\Upsilon$ is a hadron which belongs to the confinement phase of QCD whereas the quark-gluon plasma belongs to the de-confinement phase of QCD. Hence the bottomonium $\Upsilon$ cannot be formed inside the quark-gluon plasma medium but the bottomonium $\Upsilon$ is formed outside the quark-gluon plasma medium, {\it i. e.}, the bottomonium $\Upsilon$ is formed in the vacuum. This implies that the $\tau' \rightarrow \infty$ can be taken in eq. (\ref{tatp}) even if the upper limit of $\tau$ in eq. (\ref{tgf}) is $\frac{1}{T}$.

This is correct because the upper limit of $\tau=\frac{1}{T}$ in eq. (\ref{tgf}) corresponds to the fact that the in-state is equal to the out-state where the in-state is for the quarks/gluons at the finite temperature. However, the limit $\tau' \rightarrow \infty$ in eq. (\ref{tatp}) corresponds to the fact that the in-state is not equal to the out-state where the in-state is for the quarks/gluons at the finite temperature but the out-state is for the hadron which is formed in vacuum because the hadron cannot be formed inside the quark-gluon plasma. Hence the $\tau' \rightarrow \infty$ can be taken in eq. (\ref{tatp}) even if the upper limit of $\tau$ in eq. (\ref{tgf}) is $\frac{1}{T}$.

This implies that we can take  the asymptotic large time limit $\tau'\rightarrow \infty $ in eq. (\ref{tatp}) to neglect the higher energy level contributions to find
\bea
&&[\sum_{{\bf x}'} e^{i{\bf p}\cdot {\bf x}'} <in|{\cal O}_\Upsilon^\dagger(0,{\bf x}')e^{-H\tau'}{\cal O}_\Upsilon(0)|in>]_{\tau'\rightarrow \infty} = |<\Upsilon(p)|{\cal O}_\Upsilon|in>|^2 e^{-\int d\tau' E(p,\tau')}.\nonumber \\
\label{tbtp}
\eea
Since the bottomonium $\Upsilon$ is formed in vacuum we use eqs. (\ref{dtp}) and (\ref{htp}) in eq. (\ref{tbtp}) to find
\bea
&&|<\Upsilon(p)|{\cal O}_\Upsilon|in>|^2 =[\frac{\sum_{{\bf x}'} e^{i{\bf p}\cdot {\bf x}'} <in|{\cal O}_\Upsilon^\dagger(0,{\bf x}')e^{-H\tau'}{\cal O}_\Upsilon(0)|in>\times e^{\tau' E_\Upsilon(p)}}{e^{\int d\tau' [\frac{\sum_{{\bf x}''} e^{i{\bf p}\cdot {\bf x}''} <in|{\cal O}_\Upsilon^\dagger(\tau'',{\bf x}'')[\sum_{q,{\bar q},g}\int d\tau' \int d^3x'  \partial_j T^{j0}(\tau',{\bf x}')]{\cal O}_\Upsilon(0)|in>}{\sum_{{\bf x}''} e^{i{\bf p}\cdot {\bf x}''} <in|{\cal O}_\Upsilon^\dagger(\tau'',{\bf x}''){\cal O}_\Upsilon(0)|in>}]_{\tau''\rightarrow \infty}}} ]_{\tau' \rightarrow \infty}\nonumber \\
\label{titp}
\eea
which is the non-perturbative formula of the bottomonium $\Upsilon$ production amplitude from the quark-gluon plasma derived from the first principle in QCD which can be calculated by using the lattice QCD method at the finite temperature.
\section{Conclusions}
Complete suppression of the heavy quarkonium due to the screening mechanism of the quark-gluon plasma is proposed in the literature to be a signature of the quark-gluon plasma detection at RHIC and LHC. However, since the heavy quarkonium $\Upsilon$ production is experimentally measured in the Pb-Pb collisions by the ALICE collaboration at LHC one has to study the $\Upsilon$ production from the quark-gluon plasma instead of studying the $\Upsilon$ suppression due to the screening mechanism of the quark-gluon plasma. In this paper we have derived the non-perturbative formula of the $\Upsilon$ production amplitude from the quark-gluon plasma from the first principle in QCD which can be calculated by using the lattice QCD method at the finite temperature.

\end{document}